\documentclass[prd,twocolumn,floatfix,nofootinbib,amsmath,amssymb,floatfix]{revtex4}
\usepackage{graphicx,color,dcolumn,booktabs,bm,mathrsfs}
\usepackage{longtable,lscape}
\usepackage{txfonts}
\usepackage{overpic}
\usepackage{amssymb}
\usepackage{indentfirst}
\usepackage{feynmf}   
\usepackage{slashed}  
\usepackage{cases}
\usepackage{color}
\usepackage{multirow}
\usepackage{epstopdf}
\usepackage{CJK}

\usepackage[colorlinks,
            citecolor=blue,
            anchorcolor=red,
            menucolor=red,
            linkcolor=red,
            filecolor=red,
            runcolor=red,
            urlcolor=blue,
            frenchlinks=red]{hyperref}
\usepackage{subfigure}

\begin{document}

\title{The possibility of $X(4014)$ as a $D^\ast \bar{D}^\ast$ molecular state}

\author{Man-Yu Duan$^{1}$}
\author{Dian-Yong Chen$^{1,3}$}\email{chendy@seu.edu.cn}
\author{En Wang$^{2}$}\email{wangen@zzu.edu.cn}

\affiliation{$^1$School of Physics, Southeast University, Nanjing 210094, China\\
$^2$School of Physics and Microelectronics, Zhengzhou University, Zhengzhou, Henan 450001, China\\
$^3$Lanzhou Center for Theoretical Physics, Lanzhou University, Lanzhou 730000, P. R. China}

\begin{abstract}
Within the framework of the local hidden gauge approach, we have studied the near-threshold interaction of the $D^* \bar{D}^*$ channel with  quantum numbers $I(J^{PC}) = 0(0^{++})$, $0(2^{++})$, $1(0^{++})$, and $1(2^{++})$, respectively. The contact interaction is taken into account alone, since it is expected to give the dominant contribution near the threshold. One pole, which is found in the case of the quantum numbers $I(J^{PC}) = 0(0^{++})$, could be associated to the  $X(4014)$ recently observed by the Belle Collaboration. Thus, we suggest that the $X(4014)$ may be an $D^* \bar{D}^*$ molecular state with $I(J^{PC}) = 0(0^{++})$, and more precise information about $X(4014)$, such as resonance parameters and decay properties, could be useful to shed light on its structure.
\end{abstract}

\maketitle


\section{INTRODUCTION}
\label{sec:INTRODUCTION}

In last two decades, a large amount of the charmonium-like states, named $XYZ$ states, have been observed experimentally~\cite{Brambilla:2019esw,Cleven:2015era,Ma:2014zva,Liu:2019zoy,Hosaka:2016pey}. Many kinds of theoretical interpretations to their structures have been proposed, such as hadro-quarkonia~\cite{Voloshin:2007dx,Alberti:2016dru}, tetraquarks~\cite{Maiani:2014aja,Maiani:2013nmn}, hadronic molecules~\cite{Fleming:2007rp,Mehen:2011ds,Wang:2013kra,Gamermann:2009fv,Wu:2021udi,Wu:2021ezz}, kinematic effect ~\cite{Ikeno:2020mra,Nakamura:2021dix, Chen:2011pv, Chen:2013wca}, and the mixing of different components. Since most of the $XYZ$ states appear at certain hadronic threshold, hadronic molecule is one of the most promising interpretations among the various exotic states, although there are still many controversies. For instance, the hidden-charm state $X(3872)$ is quite close to the $D\bar{D}^*$ threshold~\cite{Belle:2008fma,BaBar:2007cmo}, $Z_{cs}(3985)$ is close to the $\bar{D}_s D^*/\bar{D}^*_s D$ threshold~\cite{BESIII:2020qkh}. Recently, a $T_{cc}^+$ state reported by the LHCb Collaboration has the mass very close to the $D^{*+}D^0$ threshold~\cite{LHCb:2021vvq,LHCb:2021auc}, and could well be interpreted as a $D^*D$ molecular state~\cite{Ling:2021bir,Feijoo:2021ppq,Ren:2021dsi,Chen:2021vhg,Du:2021zzh}. In addition, one loosely $D\bar{D}$ bound state with $I(J^{PC})=0(0^{++})$ predicted in Ref.~\cite{Gamermann:2006nm} is supported by the Belle and {\it BABAR} measurements of $\gamma\gamma \to D\bar{D}$ and $e^+e^-\to J/\psi D\bar{D}$~\cite{Wang:2020elp,Wang:2019evy}, and the $Z_c(4025)$ observed by the BESIII Collaboration could be interpreted as the $D^*\bar{D}^*$ molecule with $I(J^{P})=1(1^+)$~\cite{Duan:2021pll}.  As pointed out in Ref.~\cite{Dong:2020hxe}, the molecular structure should appear at any threshold of a pair of heavy-quark and heavy-antiquark hadrons with attractive interaction at threshold, and experimental information about such structures is crucial to deepen our understanding of the interactions between heavy hadrons and the internal structures of the exotic states.

In 2021, the Belle Collaboration has analyzed the two-photon process $\gamma \gamma \to \gamma \psi(2S)$ from the threshold to $4.2$~GeV for the first time~\cite{Belle:2021nuv}, and found two structures in the lineshape of the cross section. The first one with a local significance of $3.1 \sigma$ corresponds to one resonance with the mass and width as,
\begin{eqnarray}
M_1 &= & 3922.4 \pm 6.5 \pm 2.0~{\rm MeV}, \nonumber \\
\Gamma_1&=& 22 \pm 17 \pm 4~{\rm MeV},
\end{eqnarray}
which may be the $X(3915)$, $\chi_{c2}(3930)$, or the mixing of them,  and the second one with a local significance of $2.8 \sigma$ corresponds to one new resonance $X(4014)$ with the mass and width as,
\begin{eqnarray}
M_2 &= & 4014.3 \pm4.0 \pm 1.5~{\rm MeV}, \nonumber \\
\Gamma_2&=& 4 \pm 11 \pm 6~{\rm MeV}.
\label{Eq:Res}
\end{eqnarray}
As discussed in Ref.~\cite{Belle:2021nuv}, the $X(4014)$ state has a mass in agreement with the predicted mass 4012~MeV for the $J^{PC}=2^{++}$ $D^*\bar{D}^*$ molecule with the assumption of $X(3915)$ as $J^{PC}=0^{++}$ $D^*\bar{D}^*$ molecule~\cite{Guo:2013sya,Nieves:2012tt}. However, the binding energy of $D^*\bar{D}^*$ is of the order of $100$~MeV if one considers the $X(3915)$ as a $D^\ast \bar{D}$ molecular state, which is much larger than the ones of $X(3872)$, $Z_{cs}(3985)$, $T^+_{cc}$, and $Z_c(4025)$.  It should also be pointed out that the molecular explanation of $X(3915)$ is still in debate. For instance, in Refs.~\cite{Liu:2009fe,Duan:2020tsx,Kher:2018wtv} the $X(3915)$ and $\chi_{c2}(3930)$ could be assigned as the charmonia $\chi_{c0}(2P)$ and $\chi_{c2}(2P)$, respectively,  and in Ref.~\cite{Li:2015iga} $X(3915)$ is suggested as an $S$-wave $D_s^+{D}_s^-$ molecular state. Thus, it implies that the nature of the newly reported $X(4014)$ state is still unclear, and the study on the $D^*\bar{D}^*$ interaction is crucial to explore its possible internal structure.

There exist many studies about the $D^*\bar{D}^*$ interaction in literatures~\cite{Dong:2021juy,Molina:2009ct}. In our previous work, we have analysed the BESIII measurement of $e^+e^-\to (D^*\bar{D}^*)^{\pm,0} \pi^{\mp,0}$ reactions by taking into account the vector-vector  $D^* \bar{D}^*$, $K^* \bar{K}^*$, and $\rho \rho$ interactions within the framework of local hidden gauge formalism, and conclude that the $Z_c(4025)$ structure could be a shallow $D^*\bar{D}^*$ bound state~\cite{Duan:2021pll}. Recently, the interaction between charmed hadrons have been described by constant contact terms, and 229 molecular states were predicted~\cite{Dong:2021juy}. In Ref.~\cite{Dong:2021juy}, the masses of the $D^*\bar{D}^*$ bound states with $I(J^{PC})=0(0^{++})$/$0(1^{+-})$/$0(2^{++})$ are $1.82\sim 36.6$~MeV lower than the $D^*\bar{D}^*$ threshold, which favors the $D^*\bar{D}^*$ molecular explanation for the $X(4014)$ state. In the present work, we will investigate the S -wave near-threshold $D^* \bar{D}^*$ interaction within an alternative framework, i.e. the local hidden gauge approach, and the vector exchanges in the crossed channels in the hidden gauge model provide further insights into the contact interactions explored in Ref.~\cite{Dong:2021juy}. We also discuss the possibility of the X(4014) as the $D^* \bar{D}^*$ molecular state and its favored quantum numbers.

This paper is organized as follows. We will show the formalism in Sec.~\ref{sec:FORMALIISM}, and present the calculated results and related discussions in Sec.~\ref{sec:RESULTS}. Finally, Sec.~\ref{sec:Summary} will be devoted to a short summary.


\begin{figure}[t]
  \centering
 \includegraphics[scale=0.75]{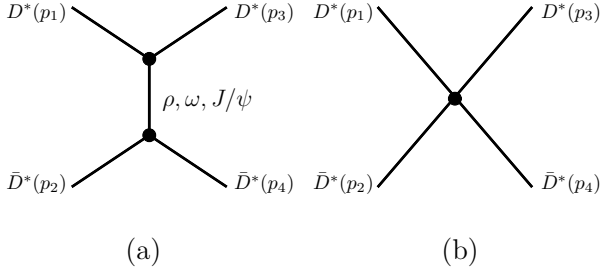}
  \caption{The sketch diagrams for $D^\ast\bar{D}^\ast\to D^\ast\bar{D}^\ast$. Diagrams (a) and (b) correspond to vector meson exchanged interaction and contact interaction, respectively, where the exchanged vector mesons can be $\rho,\ \omega$, and $J/\psi$.}
  \label{fig:mechanism}
  \end{figure}
  
  \section{FORMALISM}
\label{sec:FORMALIISM}

In this section, we investigate the $D^* \bar{D}^*$ interaction following the approach of Ref.~\cite{Molina:2009ct}. The Lagrangian is taken from the local hidden gauge formalism describing the interaction of vector mesons~\cite{Bando:1987br,Nagahiro:2008cv},
\begin{equation}
\mathcal{L}=-\frac{1}{4}\langle V_{\mu\nu}V^{\mu\nu}\rangle\ ,
\label{eq:lvv}
\end{equation}
where the symbol $\langle ~ \rangle$ stands for the trace of SU(4), and the tensor $V_{\mu\nu}$ is defined as
\begin{equation}
V_{\mu\nu}=\partial_{\mu}V_{\nu}-\partial_{\nu}V_{\mu}-ig[V_{\mu},V_{\nu}]\
\label{eq:vectensor}
\end{equation}
with $V_{\mu}$ to be
\begin{equation}
V_\mu=\left(
\begin{array}{cccc}
\frac{\omega}{\sqrt{2}}+\frac{\rho^0}{\sqrt{2}} & \rho^+ & K^{*+}&\bar{D}^{*0}\\
\rho^- &\frac{\omega}{\sqrt{2}}-\frac{\rho^0}{\sqrt{2}} & K^{*0}&D^{*-}\\
K^{*-} & \bar{K}^{*0} &\phi&D^{*-}_s\\
D^{*0}&D^{*+}&D^{*+}_s&J/\psi
\end{array}
\right)_\mu\ .
\label{eq:vfields}
\end{equation}
By expanding the effective Lagrangian in Eq. (\ref{eq:lvv}), one can obtain two types of $D^\ast \bar{D}^\ast$ interactions, which are vector-meson exchange interactions [Fig. \ref{fig:mechanism}-(a)] and contact interaction [Fig. \ref{fig:mechanism}-(b)], respectively. As for the vector meson exchange interaction, the general form of the amplitude corresponding to diagram Fig.~\ref{fig:mechanism}-(a) reads,
\begin{eqnarray}
i \mathcal{M}= ig_1 g_2 \epsilon_{1\mu}\epsilon_{2\nu} \Gamma^{\mu\alpha} \frac{-i(-g^{\nu \beta}+q^\nu q^\beta/m_{V}^2)}{q^2-m_V^2+i\epsilon} \epsilon_{3\alpha}^\ast \epsilon_{4\beta}^\ast \ .
\end{eqnarray}    
where $q$ is the momentum of the exchanged vector meson, and $\Gamma^{\mu \nu}$ is a tensor function of $g^{\mu \nu}$ and $p_i^\mu p_j^\nu$. As shown in Eq. (\ref{Eq:Res}), the mass of $X(4014)$ is very close to the threshold of $D^\ast \bar{D}^\ast$ and the width is very small. Thus, when we discuss the possibility of $X(4014)$ as a $D^\ast \bar{D}^\ast$ molecular state, we should merely consider the energy range around the threshold. In this case, the momenta of the involved mesons are very limited, i.e., $q^2\sim 0$. Then the propagator of the exchanged meson can be approximated by,
\begin{eqnarray}
	\frac{(-g^{\nu \beta}+q^\nu q^\beta/m_{V}^2)}{q^2-m_V^2+i\epsilon} \simeq \frac{g^{\nu \beta}}{m_V^2} 
\end{eqnarray}
where the terms suppressed by $\mathcal{O}(q^2/m_V^2)$ have been neglected. With the above approximation, the amplitudes of the meson exchanged diagrams are in the same form as the one of contact diagram. Thus, in the following, we can only consider the contact interaction in the present work. 

Moreover, the decay width of the molecular states are related to the contributions of the coupled channels in the present theoretical models~\cite{Zyla:2020zbs}. The  uncertainty of the widths reported by Belle Collaborations is very large, while the mass is relatively accurate. Thus as the first step, we can mainly focus on the mass of $X(4014)$, which justifies including the $D^* \bar{D}^*$ single channel in the present work. Our understanding of molecular state will greatly benefit from a systematic spectrum of hadronic molecules based on the one single channel~\cite{Dong:2021juy}. 

To depict the contact diagram, one can expand the Lagrangian in Eq. (\ref{eq:lvv}) and the effective Lagrangian for the contact term reads,
\begin{equation}
\mathcal{L}^{(c)}=\frac{g^2}{2}\langle V_{\mu}V_{\nu}V^{\mu}V^{\nu}-V_{\nu}V_{\mu}V^{\mu}V^{\nu}\rangle\ .
\label{eq:contact}
\end{equation}
With the above effective Lagrangian, one can obtain the amplitude corresponding to Fig.~\ref{fig:mechanism}-(b), which is,
\begin{eqnarray}
t^{(c)}_{D^{*+}D^{*-}\to D^{*+}D^{*-}}&=&2g^2 (\epsilon^{(1)}_\mu \epsilon^{(2)\,\mu} \epsilon^{\ast (3)}_\nu \epsilon^{\ast (4)\,\nu}+\epsilon^{(1)}_\mu \epsilon^{(2)}_\nu \epsilon^{\ast  (3)\,\mu} \epsilon^{\ast (4)\,\nu} \nonumber\\
& &-2\epsilon^{(1)}_\mu \epsilon^{(2)}_\nu \epsilon^{\ast (3)\,\nu} \epsilon^{\ast (4)\,\mu} )\ ,\nonumber\\
t^{(c)}_{D^{*+}{D}^{*-} \to D^{*0}\bar{D}^{*0}}&=&g^2 (\epsilon^{(1)}_\mu \epsilon^{(2)\,\mu} \epsilon^{\ast (3)}_\nu \epsilon^{\ast (4)\,\nu} +\epsilon^{(1)}_\mu \epsilon^{(2)}_\nu \epsilon^{\ast (3)\,\mu} \epsilon^{\ast (4)\,\nu} \nonumber\\
& &-2\epsilon^{(1)}_\mu \epsilon^{(2)}_\nu \epsilon^{\ast (3)\,\nu} \epsilon^{\ast (4)\,\mu} )\ ,
\label{eq:DD}
\end{eqnarray}
where the indices 1, 2, 3, and 4 correspond to the particles with the momenta $p_1$, $p_2$, $p_3$, and $p_4$ in Fig.~\ref{fig:mechanism}-(b). By considering the spin projection operators and the isospin doublets charmed and anticharmed mesons as Refs.~\cite{Molina:2009ct,Molina:2008jw}, we obtain the kernels of $D^*\bar{D}^*$ interactions for different isospin and total angular momentum, which are,
\begin{eqnarray}
V^{I = 0,J = 0}_{D^*\bar{D}^*\to D^*\bar{D}^*}&=&6g_{D}^2,   \qquad    V^{I = 0,J = 2}_{D^*\bar{D}^*\to D^*\bar{D}^*}=-3g_{D}^2, \nonumber\\
V^{I = 1,J = 0}_{D^*\bar{D}^*\to D^*\bar{D}^*}&=&2g_{D}^2,    \qquad   V^{I = 1,J = 2}_{D^*\bar{D}^*\to D^*\bar{D}^*}=-g_{D}^2,
\label{eq:potential}
\end{eqnarray}

With the above potentials, one can search the possible poles by solving the Bethe-Salpeter equation, which is
\begin{equation}
T=[1-VG]^{-1}V\ ,
\label{eq:BS}
\end{equation}
where the $G$ is the two-meson loop function given by
\begin{equation}
G=i\int\frac{d^4q}{(2\pi)^4}\frac{1}{q^2-m_1^2+i\epsilon}\frac{1}{(q-P)^2-m_2^2+i\epsilon}\ ,
\label{eq:loopex}
\end{equation}
with $m_1$ and $m_2$ the masses of the two mesons. $q$ is the four-momentum of the meson in the centre of mass frame, and $P$ is the total four-momentum of the meson-meson system. 

In the present work, we use the dimensional regularization method as indicated in  Refs.~\cite{Molina:2009ct,Duan:2020vye}, and in this scheme, the two-meson loop function $G$ can be expressed as,
\begin{eqnarray}
G&=&\frac{1}{16\pi^2}\left[\alpha+\log\frac{m_1^2}{\mu^2}+\frac{m_2^2-m_1^2+s}{2s}\log\frac{m_2^2}{m_1^2}\right. \nonumber\\
&&+\frac{|\vec{q}\,|}{\sqrt{s}}\left(\log\frac{s-m_2^2+m_1^2+2|\vec{q}\,|\sqrt{s}}{-s+m_2^2-m_1^2+2|\vec{q}\,|\sqrt{s}}\right. \nonumber \\
&&+\left. \left. \log\frac{s+m_2^2-m_1^2+2|\vec{q}\,|\sqrt{s}}{-s-m_2^2+m_1^2+2|\vec{q}\,|\sqrt{s}}\right)\right] ,
\label{eq:loopexdm}
\end{eqnarray}
where $\mu$ and $\alpha$ are model parameter, while $\vec{q}\,$ is the momentum of the meson in the centre of mass frame, which reads,
\begin{equation}
|\vec{q}\,|=\frac{\sqrt{\left[s-(m_1+m_2)^2\right]\left[s-(m_1-m_2)^2\right]}}{2\sqrt{s}} .
\end{equation}

In the complex plane for a general $\sqrt{s}$, the loop function in the second Riemann sheet can be written as
\begin{equation}
\label{eq:lloopexdm}
G^{II}(\sqrt{s})=G^{I}(\sqrt{s})+i\frac{|\vec{q}\,|}{4\pi\sqrt{s}},\qquad {\rm Im}(|\vec{q}\,|)>0,
\end{equation}
where $G^{II}$ refers to the loop function in the second Riemann sheet, and $G^{I}$ is the one in the first Riemann sheet as given by Eq.~\eqref{eq:loopexdm} for the $D^* \bar{D}^*$ channel. When searching for the poles, we use $G^{I}$ for Re$(\sqrt{s}) < m_1+m_2$, and use $G^{II}$ for Re$(\sqrt{s}) > m_1+m_2$. 


\section{RESULTS AND DISCUSSIONS}
\label{sec:RESULTS}

Before we discuss the numerical results, the values of the relevant parameters should be clarified, which include the coupling constants $g_D$, the parameter $\mu$, and the subtraction constant $\alpha$. As for $\alpha$, it is usually fixed to be $ \mu= 1$ ~GeV as indicated in Ref.~\cite{Molina:2009ct}. In general, the coupling constant $g_D=m_{D^\ast}/(2f_D)=6.9$~\cite{Molina:2009ct}, however, in the present work, the vector meson exchanged interactions have been absorbed by the contact diagrams, thus, the value of $g_D$ will be changed in the present estimation. Here we vary $g_D$ from 5 to 8 to check the parameter dependence of the results. It should be clarified that the value of the subtraction constant $\alpha$ can not be estimated by the first principle in the present frame and usually be considered as a phenomenological model parameter~\cite{Gamermann:2006nm,Duan:2021pll}, which can be fixed by reproducing some relevant experimental measurements. For example, in Ref.~\cite{Molina:2009ct} the value of $\alpha$ is fixed to be -2.07 in order to get a pole around 3940~MeV in $I=0,\ J=0$ channel, which is associated to the $X(3915)$. In the present work, we mainly focus on the mass of the $X(4014)$, however, the experimental data for $\gamma \gamma \to \gamma \psi(2S)$ are very inaccurate around $4014$~MeV~\cite{Belle:2021nuv}, thus, it is difficult to fit our results with experimental data. Here, We will vary the subtraction constant $\alpha$ in a large range to check whether one can reproduce the mass of $X(4014)$ in a reasonable parameter range. 

 \begin{figure}[t]
    \centering
    \includegraphics[scale=0.7]{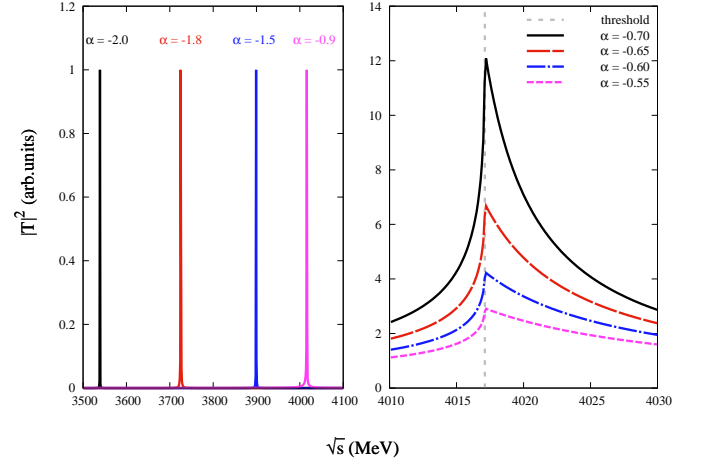}
    \caption{The modulus squared of the amplitude $|T_{D^* \bar{D}^* \to D^* \bar{D}^*}|^2$ for the case of $I=0, \ J=0$ depending on the parameter $\alpha$.}
    \label{fig:pole-threshold}
  \end{figure}
	
With the above formalisms, we can  estimate the amplitudes for the $D^*\bar{D}^*\to D^*\bar{D}^*$ transition with quantum numbers $I(J^P)=0(0^+)$, $0(2^+)$, $1(0^+)$, $0(2^+)$. Here, we take the case of $I=0$, $J=0$ as an example to show the modulus squared of the amplitude $|T_{D^\ast \bar{D}^\ast \to D^\ast \bar{D}^\ast}|$ depending on the subtraction constant $\alpha$ in Fig.~\ref{fig:pole-threshold}, where the coupling constants $g_D=6.9$. From the figure, one can find when $\alpha$ is very small, such as $\alpha=-2.0$, there is a pole around 3540 MeV, by increasing $\alpha$, the pole moves to the threshold, then the modulus squared of the amplitude behaves like a cusp when $\alpha$ is greater than -0.7, which will not correspond to any bound state any more.

  \begin{figure}[htb]
    \centering
    \includegraphics[scale=1]{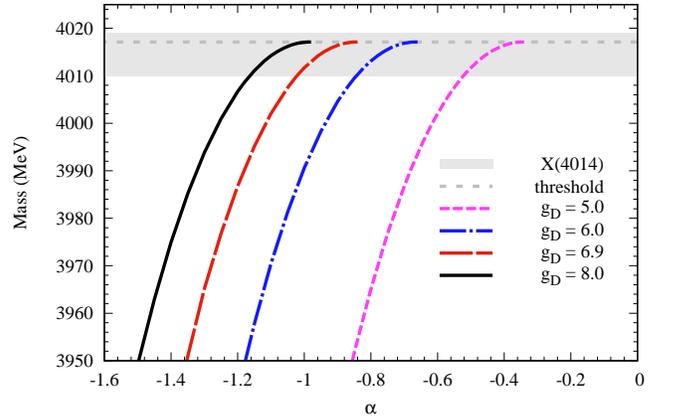}
    \caption{The pole mass depending on the parameter $\alpha$ for the case of $I = 0$, $J = 0$. The grey horizontal line and  band indicates the $D^\ast \bar{D}^\ast$ threshold and the mass range of $X(4014)$ measured by Belle collaboration~~\cite{Belle:2021nuv}.}
    \label{fig:mass-00}
  \end{figure}
  
\begin{table}[htb]
\caption{The parameter range determined by reproducing the measured mass of $X(4014)$ with different $I(J^{PC})$ assignments. \label{Tab:alpharange} }
\begin{tabular}{p{1.cm}<\centering p{2cm}<\centering  p{4cm}<\centering }
\toprule[1pt]
$I(J^{PC})$ & $g_D$ & $\alpha$ Range \\
\midrule[1pt]
\multirow{4}{*}{$0(0^{++})$} & 5.0 & $-0.54 \sim -0.34 $ \\
 & 6.0 & $-0.86 \sim -0.66 $ \\
 & 6.9 & $-1.04 \sim -0.84 $ \\
 & 8.0 & $-1.18 \sim -0.98 $ \\
\midrule[1pt]
$1(0^{++})$& 8.0  & $-0.36 \sim -0.16 $ \\
\bottomrule[1pt]   	
\end{tabular}	
\end{table}

In Fig.~\ref{fig:mass-00}, we present the estimated pole  for the case of $I=0, \ J=0$ depending on the coupling constants $g_D$ and the subtraction constants $\alpha$. Here, the coupling constants $g_D$ are set to be 5.0, 6.0, 6.9, and 8.0, respectively. From the figure, one can find that, for a certain $g_D$, the pole mass increases with the increasing of $\alpha$, and it appears the overlaps between the theoretical estimation and Belle data. In Table \ref{Tab:alpharange}, we collect the parameter range determined by reproducing the measured mass of $X(4014)$, where one can find for different $g_D$,  the mass of $X(4014)$ could be reproduced by adjusting the subtraction parameter $\alpha$.

\begin{figure}[htb]
    \centering
    \includegraphics[scale=0.95]{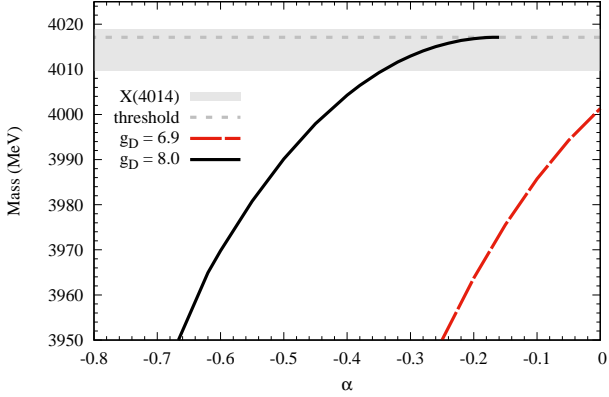}
    \caption{The same as Fig.~\ref{fig:mass-00} but for the case of $I = 1$, $J = 0$.}
    \label{fig:mass-10}
  \end{figure}

\begin{figure}[htb]
    \centering
    \includegraphics[scale=0.65]{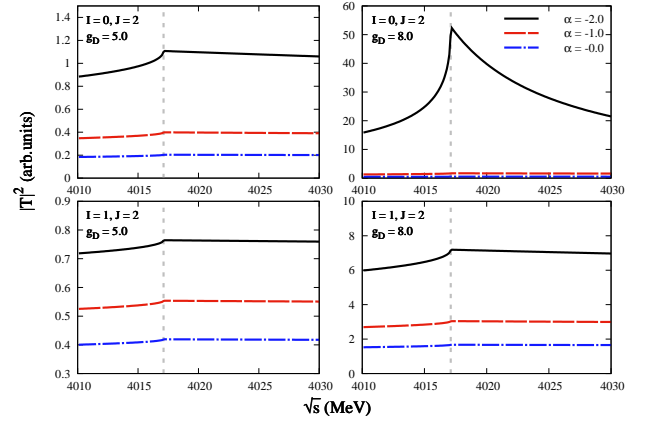}
    \caption{The same as Fig.~\ref{fig:pole-threshold} but for the cases of $I = 0$, $J = 2$ and  $I = 1$, $J = 2$.}
    \label{fig:J2}
  \end{figure}

In Fig.~\ref{fig:mass-10}, we present the estimated pole mass for the case of $I=1$, $J=0$ depending on the subtraction constant $\alpha$. Similar to the case of $I=0$, $J=0$, the pole mass increases with the increasing of $\alpha$ for a certain $g_D$. However, our estimations indicate that when we take $g_D=6.9$, the pole mass always far below the measured mass of $X(4014)$ even when $\alpha=0$.  When we take a larger $g_D$, for example $g_D=8.0$, we can reproduce Belle data with $-0.36<\alpha <-0.16$. Besides the cases of $J=0$, we also check the cases of $J=2$ to further evaluate the favored quantum numbers of $X(4014)$. In Fig.~\ref{fig:J2}, we present modulus square of the amplitude for the cases of $I=0$, $J=2$ and $I=1$, $J=2$ depending on the parameter $\alpha$. From the figure one can find in the considered parameter space, i.e., $5.0 <g_D<8.0$, $-2.0<\alpha<0$, there is no pole in the modulus square of the amplitudes, which indicates that the $D^\ast \bar{D}^\ast$ can not form a bound state with $J=2$ based on our calculations.

\section{Summary}
\label{sec:Summary}

Inspired by recent observation of $X(4014)$ in the process $\gamma \gamma \to \gamma \psi(2S)$, we have studied the hidden-charm $D^* \bar{D}^*$ interaction within the framework of the local hidden gauge approach, where the possible $I(J^{PC})$ quantum numbers could be $0(0^{++}), 0(2^{++}), 1(0^{++})$, and $1(2^{++})$ respectively. For the case of $I(J^{PC})=0(0^{++})$, our estimations indicate in a large coupling constant range, i.e., $5.0 <g_D<8.0$, one can reproduce the measured mass of $X(4014)$ by adjusting the subtraction parameter $\alpha$. As for the case of $I(J^{PC})=1(0^{++})$, one can also reproduce the measured mass but only with a large coupling constants $g_D$. As for the cases of $I(J^{PC})=0(2^{++})$ and $I(J^{PC})=1(2^{++})$, our calculations show that there is no pole in the modulus square of the amplitudes in the considered parameter space.

To summarize, in the $D^\ast \bar{D}^\ast$ molecular scenario, our estimations weakly favor the $I(J^{PC})=0(0^{++})$ assignment for $X(4014)$ comparing to $I(J^{PC})=1(0^{++})$, and disfavor the  $J=2$ assignments. However, it should be clarified that the local significance of $X(4014)$ is only 2.8$\sigma$, more experimental information about this state is crucial to shed light on its nature.

Before the end of this work, we should mention that there already exist  some hints of $X(4140)$ in the experimental data. For instance, the Belle Collaboration has measured the inclusive process $e^+e^-\to J/\psi X$, and one can observe a peak around 4014~MeV in the $M_{\rm rec}(J/\psi)$ distribution, as shown in Fig.~1 of Ref.~\cite{Belle:2005lik}. Later, the Belle Collaboration reported the study of the processes $e^+e^-\to J/\psi D^{(*)}\bar{D}^{(*)}$, and there is more events in the energy region $4000<M_{D\bar{D}^*}<4025$~MeV of the $D\bar{D}^*$ invariant mass distribution, as depicted in Fig.~2(b) of Ref.~\cite{Belle:2007woe}. In addition, in the cross sections of the process $\gamma\gamma \to J/\psi \omega$ measured by the Belle Collaboration, one can find the events concentrate in the region $4010\sim 4020$~MeV~\cite{Belle:2009and}. All the above hints imply that one state with a mass close to the $D^*\bar{D}^*$ threshold and a narrow width exists, which coincides with the $X(4140)$ observed by the Belle Collaboration~\cite{Belle:2021nuv}.

\section*{Acknowledgement}
The authors thank Prof. Zhi-Hui Guo for valuable discussion and suggestive comments. This work is supported by the National Natural Science Foundation of China under Grant No.11775050, 12175037. This work is also supported by the Natural Science Foundation of Henan under Grand No. 222300420554,  the Project of Youth Backbone Teachers of Colleges and Universities of Henan Province (2020GGJS017), the Youth Talent Support Project of Henan (2021HYTP002), and the Open Project of Guangxi Key Laboratory of Nuclear Physics and Nuclear Technology, No.NLK2021-08.


\end{document}